\documentclass[prl,aps,twocolumn,floats,showpacspsfig]{revtex4}
\usepackage{amssymb}

\usepackage{epsfig}

\newcommand{\be}{\begin{equation}}
\newcommand{\ee}{\end{equation}}
\newcommand{\bea}{\begin{eqnarray}}
\newcommand{\eea}{\end{eqnarray}}

\newcommand{\p}{\partial}
\newcommand{\s}{\sigma}

\newcommand{\la}{\langle}
\newcommand{\ra}{\rangle}
\newcommand{\rd}{\mbox{d}}
\newcommand{\ri}{\mbox{i}}

\usepackage{epstopdf}
\begin{document}
\title{Spin density wave formation  in graphene \\facilitated by the in-plane magnetic field}
\author{ D. E. Kharzeev$^a$, S. A. Reyes$^b$ and A.  M. Tsvelik$^b$}
\affiliation{
a) Department of Physics,
Brookhaven National Laboratory, Upton, NY 11973-5000, USA;\\
b) Department of Condensed Matter Physics and Materials Science, Brookhaven National Laboratory, Upton, NY 11973-5000, USA and
Department of Physics and Astronomy, Stony Brook University, Stony Brook, NY 11794-3800, USA}

\date{\today}

\begin{abstract}
  We suggest that by applying a magnetic field lying in the plane of graphene layer one may facilitate an excitonic condensation of electron-hole pairs with opposite spins and chiralities. The provided calculations  yield a conservative estimate for the transition temperature $T_c \sim 0.1~ B$. 

\end{abstract}

\pacs{ 71.10.Pm, 72.80.Sk}
\maketitle
\narrowtext

\sloppy

  Charge Density Wave (CDW) and Spin Density Wave (SDW) formation in graphene is an intriguing possibility which has been discussed in the literature \cite{khv},\cite{gus},\cite{herbut}. Even more so since the reduced dimensionality of   graphene  will strongly affect the character of the corresponding ordered state probably making it critical. It was also pointed out that magnetic field  perpendicular to the graphene layers may facilitate formation of the CDW \cite{khv},\cite{gus}. The CDW order parameter discussed in these papers establishes  different population densities on the two sublattices of graphene (we will call them $u$ and $v$ ones). This corresponds to a site-centered CDW. In this paper we suggest that by applying a magnetic field in the graphene plane or a Weiss exchange field \cite{matros} one can facilitate a formation of a bond-centered  spin density wave. The corresponding order parameter has U(1)$\times$Z$_2$ symmetry and combines  fermion operators with  opposite spins and chiralities. 

 The tight binding Hamiltonian is 
\bea
H_0 = - t\sum_{{\bf r},i\s}u^+_{\s}({\bf r})v({\bf r} + {\bf b}_i) + H.c.
\eea
where ${\bf r}$ belong to a triangular lattice (see Fig. 1).
 
\begin{figure}
[ht]
\begin{center}
\epsfxsize=0.35\textwidth
\epsfbox{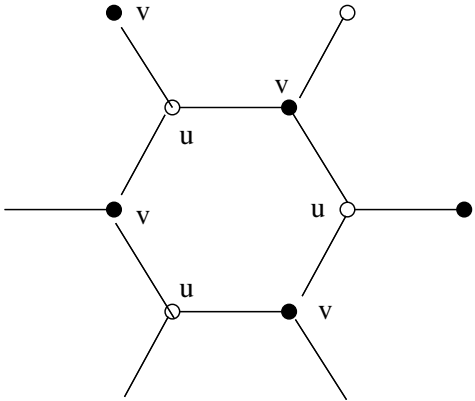}
\end{center}
\caption{Crystalline lattice of graphene. }
\label{uv}
\end{figure}
Tunneling on next nearest neighbours generates the so-called trigonal warping of the spectrum. We will not discuss this  small deformation  of the spectrum since it does not destroy the nesting between the bands of different chirality and therefore will not have any adverse effect on the phenomena discussed in this paper. 

The Hamiltonian describing states close to the tips of the Dirac cones is 
\bea
&& \omega \hat I - \hat H_0 = \\
&& \sum_{\s = \pm 1}\Psi_{\s}^+\left(
\begin{array}{cccc}
\omega - \s B & vk & 0 &0\\
v\bar k &\omega - \s B & 0 &0\\
0 &0 & \omega + \s B & - v\bar k\\
0 & 0& - vk & \omega + \s B
\end{array}
\right)\Psi_{\s}, \nonumber\\
 && \Psi^+ = \left(
u^+_{\s}({\bf k} + {\bf Q}), 
v^+_{\s}({\bf k} +  {\bf Q}), 
u^+_{-\s}({\bf k}-{\bf Q}), 
v^+_{-\s}({\bf k} -{\bf Q})
\right)\nonumber
\eea
Here $u,v$ operators annihilate electrons on different sublattices and $k = k_x + \ri k_y$, ${\bf Q} = (1,1/\sqrt 3)2\pi/a\sqrt 3$. Since magnetic field $B$ 
lies  in the plane, it affects only spin (the same effect can probably be achieved by the exchange field when one brings  graphene sample in a close proximity to a ferromagnet). The in-plane magnetic field leads to Zeeman splitting of the bands (see Fig. 2). 
\begin{figure}
[ht]
\begin{center}
\epsfxsize=0.5\textwidth
\epsfbox{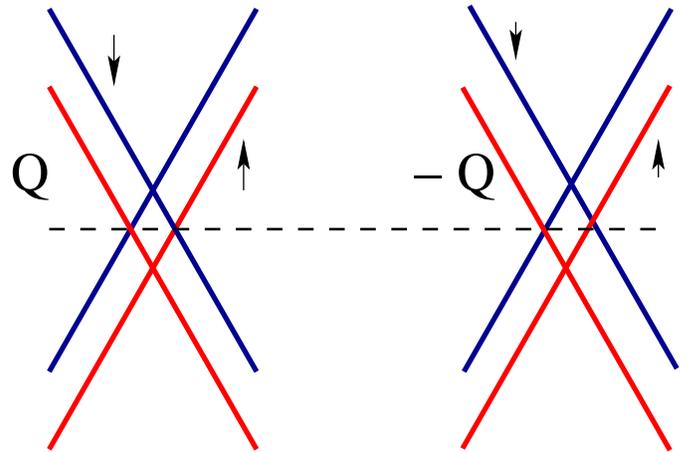}
\end{center}
\caption{Graphene bands split by the in-plane magnetic field. }
\label{bands}
\end{figure}
The predominant interaction is the  Coulomb one: 
\begin{widetext} 
\bea
&& V =  \frac{1}{2}\sum_{k,k',q;\tau = \pm 1}[u^+_{\s}(k' + q + \tau Q)u_{\s}(k + q +\tau Q) + v^+_{\s}(k' + q +\tau Q)v_{\s}(k + q + \tau Q)]\frac{2\pi e^2}{|{\bf k} - {\bf k}'|}\nonumber\\
&& \times[u^+_{\s'}(k + \tau' Q)u_{\s'}(k' + \tau' Q) + v^+_{\s'}(k  +\tau' Q)v_{\s'}(k' + \tau' Q)] \label{int}
\eea 
\end{widetext}
 The bare Coulomb interaction in graphene is fairly strong $e^2/\hbar v \approx 2$, but it is screened:
\bea
V(q) = \frac{2\pi e^2}{|q| - 2\pi e^2 N\Pi(\omega,q)}
\eea
 where $N=4$ is the number of fermion species (valley and spin). Assuming that the self-consistent interaction is weak we can get an estimate of this interaction by substituting there  the polarization loop $\Pi(\omega,q)$ for bare electrons. Then the effective interaction can be  approximated as
\bea
V(q) = \frac{1}{N}[\rho(\epsilon_F)]^{-1}\left\{\begin{array}{c} 1, ~~ |q| < B\\B/C|q|, ~~ |q| > B
\end{array}
\right\}
\eea
where $\rho(\epsilon_F) =k_F/2\pi$ and $C(N) \sim 1$ is constant (according to \cite{khv} $C(\infty) = \pi/4$).

 The energetics selects order parameters where electrons pair with holes of the opposite spin and chirality. There are two complex  order parameters:
\bea
&& \Delta^{\s}_{14}(k) = \sum_{{\bf k}'}\la u_{\s}({\bf k}'+{\bf Q})V(|{\bf k} - {\bf k}'|)v^+_{-\s}({\bf k}'-{\bf Q})\ra, \nonumber\\
&&  \Delta^{\s}_{23}(k) = \sum_{{\bf k}'}\la v_{-\s}({\bf k}'+{\bf Q})V(|{\bf k} - {\bf k}'|)u^+_{\s}({\bf k}'-{\bf Q})\ra \label{delta}
\eea
As we shall see, on the mean field level these order parameters have equal amplitudes independent of spin:  $\Delta^{\s}_{14}(k) = \Delta^{\s'}_{23}(k) = \Delta(k)$.
 Plugging (\ref{delta}) into (\ref{int}) we obtain the mean field Hamiltonian:
\bea
&&\omega \hat I - \hat H_{MF} = \\
&& \sum_{\s=\pm 1}\Psi^+_{\s}\left(
\begin{array}{cccc}
\omega - \s B & vk & 0 &\Delta(k)\\
v\bar k &\omega -  \s B & \Delta(k) &0\\
0 &\Delta(k) & \omega + \s B & - v\bar k\\
 \Delta(k)& 0& - vk & \omega  + \s B
\end{array}
\right)\Psi_{\s} \nonumber
\eea
The spectrum is 
\be
E^2_{\pm} = |\Delta(k)|^2 + v^2(|k| \pm B)^2
\ee

The self-consistency conditions are 
\bea
&& \Delta_{14}(k) = \\
&& \int \frac{\rd^2 p}{(2\pi)^2}V(|{\bf k} - {\bf p}|)\left[\Pi_{11,44}(p) \Delta_{14}(p) + \Pi_{12,34}(p) \Delta_{23}(p)\right]\nonumber\\
&& \Delta_{23}(k) = \\
&& \int \frac{\rd^2 p}{(2\pi)^2}V(|{\bf k} - {\bf p}|)\left[\Pi_{22,33}(p) \Delta_{14}(p) + \Pi_{21,43}(p) \Delta_{23}(p)\right]\nonumber
\eea
where 
\bea
&& \Pi_{11,44}(p) = \Pi_{22,33}(p) = T\sum_n G_{11}^{\uparrow}(\omega_n,p)G_{44}^{\downarrow}(\omega_n,p)\nonumber\\
&& \Pi_{12,34}(p) = \Pi_{21,43}(p)= T\sum_n G_{12}^{\uparrow}(\omega_n,p)G_{34}^{\downarrow}(\omega_n,p) \nonumber
\eea
As we said, these equations yield $\Delta_{14}(k) = \Delta_{23}(k) = \Delta(k)$. Summing over the Matsubara frequencies and integrating over angles we get the following equation:
\bea
&& \Delta(p) = \int \frac{\rd^2 k}{(2\pi)^2}\frac{V({\bf p} - {\bf k})}{2}\times\\
&& \left\{\frac{\tanh[(|k| + B)\beta/2]}{|k| + B} + \frac{\tanh[(|k| - B)\beta/2]}{(|k| - B)}\right\}\Delta(k)\nonumber
\eea
In integrating over angles we will adopt the same approximation as in \cite{khv} and will replace the resulting elliptic function $K(k/p)$ by 1. In the limit of $N \rightarrow \infty$ these equations yield $T_c \sim B\exp(-N)$. We believe that using this formula for  $N=4$ substantially underestimates $T_c$. The approximate equations for the mean field critical temperature $\beta^* = B/2T_c >> 1$ is (here $x = p/B$):
\begin{widetext}
\bea
\Delta(x) = \frac{1}{2N}\int_0^{\infty}\frac{y\rd y}{C \mbox{max}(x,y) + 1}\left\{\frac{\tanh[(y+1)\beta^*]}{y+1} + \frac{\tanh[(y-1)\beta^*]}{y-1}\right\}\Delta(y)\label{self3}
\eea
\end{widetext}
We found a good interpolation formula will split the interval of integration into (0,A) and $(A, \infty)$ where $A > 1$. At $x < A$ the interaction will be approximated as constant, and at $x > A$ as $(Cx)^{-1}$. We look for the solution of this equation as
\bea
\Delta(x) = \frac{\Delta_0}{(x + 1)^{\gamma}}
\eea 
where  $\gamma = 1/2 + \sqrt{1/4 - 1/CN}$ 
and 
\be
T_c \approx B\exp[- (N - 1/\gamma)]
\ee
Assuming that $C(4) \approx 1$ we get $\gamma(4) \approx 1/2$ and the estimate $T_c \sim 10^{-1}B$.
  

  Few words about the effects of reduced dimensionality. As is well known, in two dimensions  continuos U(1) symmetry cannot be spontaneously broken. In this case the second order phase transition is replaced by the Berezinskii-Kosterlitz-Thouless (BKT) one and the low temperature phase is critical. The BKT termperature is determined by the equation 
\bea
T_{BKT} = \frac{\pi}{2}\rho(T_{BKT})
\eea
where $\rho(T)$ is the temperature dependent stiffness. $\rho(T)$ vanishes at $T_c$ and reaches some constant value at $T=0$. Since the system is not strongly interacting, $\rho(0) >> T_c$.  Therefore the BKT transition temperature occurs close to the mean field and one should expect to see the effects of fluctuations only in a fairly narrow region around the transition temperature. 

 Our estimate of the transition temperature indicates that it is not too low and for realistic magnetic fields $B \sim 10-20~ {\rm T}$ probably lies in the interval $1~$ K or even larger. In a hypothetical construction a graphene layer is brought into a microscopic contact with a ferromagnetic film; in this case the Weiss fields can probably be made higher by an order of magnitude.
 
This work was  supported  by 
the DOE under contract No. DE-AC02 -98 CH 10886.  We acknowledge inspirational conversations with I. Zaliznyak. AMT also acknowledges valuable discussions with L. Levitov and D. Khveshchenko.


\begin{thebibliography}{99}


\bibitem{khv}D. V. Khveshchenko, Phys. Rev. Lett. {\bf 87}, 246802 (2001); ibid. {\bf 87}, 206401 (2001); D. V. Khveshchenko and H. Leal, Nucl. Phys. B{\bf 687}, 323 (2004);  D. V. Khveshchenko and W. F. Shively, Phys. Rev. B{\bf 73}, 115104 (2006). 
 \bibitem{gus} E. V. Gorbar, V. P. Gusynin, V. A. Miransky and I. A. Shovkovy, Phys. Rev. B{\bf 66}, 045108 (2002).
 \bibitem{matros} I. Zaliznyak, private communication.
\bibitem{herbut} I. F. Herbut, cond-mat/0606195.


\end{thebibliography}
\end{document}